\documentclass[manuscript]{aastex}

\usepackage{epsfig}

\def\'#1{\ifx#1i{\accent"13\i}\else{\accent"13#1}\fi}

\newcommand{\kms}{{\rm ~km~s}^{-1}}
\newcommand{\dnob} {\Delta N_{\rm OB}}
\newcommand{\fL} {f_{\rm L}}

\newcommand{\Mc} {M_{\rm C}}

\newcommand{\MI} {M_{\rm I}}

\newcommand{\Mmax} {M_{\rm max}}

\newcommand{\Ms} {M_{\rm S}}
\newcommand{\Mrms} {\mathcal{M}_{\rm rms}}
\newcommand{\msol}{{\rm M}_{\odot}}

\newcommand{\Msun}{{\rm M}_{\odot}}
\newcommand{\MtI} {M_{\rm tI}}
\newcommand{\MtII} {M_{\rm tII}}
\newcommand{\MtIII} {M_{\rm tIII}}

\newcommand{\ninf}{n_{\rm inf}}

\newcommand{\nsf} {n_{\rm SF}}

\newcommand{\pcc}{{\rm ~cm}^{-3}}

\newcommand{\Rc} {R_{\rm C}}
\newcommand{\Rinf}{R_{\rm inf}}

\newcommand{\SHI}{\Sigma_{\rm HI}}
\newcommand{\SHtwo}{\Sigma_{{\rm H}_2}}
\newcommand{\Sg}{\Sigma_{\rm gas}}

\newcommand{\SSFR}{\Sigma_{\rm SFR}}

\newcommand{\tff} {{t_{\rm ff}}}
\newcommand{\tob} {t_{\rm OB}}

\newcommand{\vinf} {v_{\rm inf}}
\newcommand{\VS}{V\'azquez-Semadeni}
\newcommand{\xob} {x_{\rm OB}}

\newcommand{\etal}{et al.}
\newcommand{\beq}{\begin{equation}}
\newcommand{\eeq}{\end{equation}}
\newcommand{\bce}{\begin{center}}
\newcommand{\ece}{\end{center}}

\slugcomment{Submitted to ApJ}

\shorttitle{Evolutionary Model for Collapsing Molecular Clouds}
\shortauthors{Zamora-Avil\'es et al.}

\begin{document}


\title{An Evolutionary Model for Collapsing Molecular Clouds and Their
Star Formation Activity}


\author{Manuel Zamora-Avil\'es\altaffilmark{1}, Enrique
V\'azquez-Semadeni\altaffilmark{1} and Pedro Col\'in\altaffilmark{1}}


\altaffiltext{1}{Centro de Radioastronom\'\i a y Astrof\'\i sica,
Universidad Nacional Aut\'onoma de M\'exico, Apdo. Postal 3-72, Morelia,
Michoac\'an, 58089, M\'exico}


\begin{abstract}

We present an idealized, semi-empirical model for the evolution of
gravitationally contracting molecular clouds (MCs) and their star
formation rate (SFR) and efficiency (SFE). The model assumes that the
instantaneous SFR is given by the mass above a certain density threshold
divided by its free-fall time. The instantaneous number of massive stars
is computed assuming a Kroupa IMF. These stars feed back on the cloud
through ionizing radiation, eroding it. The main controlling parameter
of the evolution turns out to be the maximum cloud mass, $\Mmax$. This
allows us to compare various properties of the model clouds against
their observational counterparts. A giant molecular cloud (GMC) model ($\Mmax
\sim 10^5~\Msun$) adheres very well to the evolutionary scenario
recently inferred by Kawamura et al. (2009) for GMCs in the Large Magellanic Cloud. 
A model cloud with $\Mmax \approx 2000~\Msun$ evolves in the Kennicutt-Schmidt
diagram first passing through the locus of typical low- to-intermediate
mass star-forming clouds, and then moving towards the locus of high-mass
star-forming ones over the course of $\sim 10$ Myr. Also, the stellar
age histograms for this cloud a few Myr before its destruction agree
very well with those observed in the $\rho$-Oph stellar association,
whose parent cloud has a similar mass, and imply that the SFR of the
clouds increases with time. Our model thus agrees well with various
observed properties of star-forming MCs, suggesting that the scenario of
gravitationally collapsing MCs, with their SFR regulated by stellar
feedback, is entirely feasible and in agreement with key observed
properties of molecular clouds.

\end{abstract}



\section{Introduction} \label{sec:intro}

A crucial ingredient in understanding the star formation efficiency
(SFE) of giant molecular clouds (GMC) is the study of their evolution,
from their formation to their destruction by the massive stars they
form. A still unsolved problem is whether the GMCs are in approximate
virial equilibrium, or rather they are in gravitational contraction.
Initially, Goldreich \& Kwan (1974) proposed that the supersonic
linewidths observed in GMCs correspond to global gravitational
contraction, but Zuckerman \& Palmer (1974) readily argued that if all
the molecular gas in the Galaxy were in free fall, then the total star
formation rate (SFR) in the Galaxy would be about two orders of
magnitude higher than observed (we will refer to this as the ``SFR
conundrum''). Zuckerman \& Evans (1974) subsequently suggested that the
linewidths could correspond to small-scale turbulent motions, giving
rise to the notion that clouds are quasi-equilibrium entities, a notion
that has survived until today \citep[see e.g., the reviews by][]
{MacLow-Klessen04, MO07}.

Since then, most theoretical models of star formation (SF) have been
based on the assumption 
that turbulence provides support against the clouds' self-gravity,
and allows them to maintain a quasi-virial equilibrium state, thus
preventing global collapse and maintaining a low global SFR 
\citep[e.g.,][]{NS80, McKee89, Matzner02, KM05, LN06, NL07,
Wang_etal10}. In these models, the 
turbulence is maintained by the energy feedback into the cloud from the
stars it forms. One interesting
model where strict equilibrium was not assumed, was that by
\citet{KMM06}, where the 
fully time-dependent Virial Theorem was solved numerically for a cloud
under the influence of its self-gravity and the pressure produced by
feedback from H II regions, although the cloud was restricted to have a
spherical geometry, and mass loss by the cloud due to ionizing radiation
by massive stars was not considered. Those authors found that clouds
undergo a few expansion-contraction oscillations, until they are finally
dispersed, and the SFEs over the clouds' lifetimes were found to be
$\sim 5$ -- 10\%. More recently, a similar model, with the same
restrictions but including mass accretion from the environment, was
considered by \citet{Goldbaum+11}. In this model, the clouds again reach
virial equilibrium, and maintain roughly constant column densities.

However, recent theoretical and observational evidence has suggested a
return to the global gravitational contraction scenario of Goldreich \&
Kwan (1974) \citep[e.g.,][]{HBB01, BH04, HB07,
PHA07, Galvan+09,VS+07,VS+09,Csengeri+10,Schneider+10}. Additionally,
\citet{CB05} have suggested that molecular cloud turbulence, rather than
directly producing Jeans-unstable clumps, only produces the seed
nonlinear density fluctuations for subsequent gravitational
fragmentation, which proceeds on different timescales
due to the spatial variations on the local free-fall
time induced by the turbulence \citep{HH08}. But then, if we again allow for
clouds and their substructures to be in gravitational contraction, it
is necessary to find a solution for the SFR conundrum in this scenario. The
semi-empirical model presented here investigates whether stellar
feedback can accomplish this.

Our model is motivated by the numerical simulations of
\citet{VS+10}, who have investigated the evolution of clouds formed by
the collision of warm neutral medium (WNM) cylindrical streams, including stellar feedback
from ionization heating from massive stars. Those authors found that the
clouds are in general not stabilized by the feedback, but rather are
either dispersed or continue to contract globally, depending on their mass.
Here, we construct a model that attempts to capture the phenomenology
observed in those simulations, scanning the space of the
parameters that determine the physical properties of the cloud. 
\citet{Yetli+10} have also presented a parameter-space study, but using
numerical simulations, and without feedback, while
\citet{Dib+11} have produced an analytical model similar to ours, but
aimed at investigating the effect of varying metallicity on the SFE.

The plan of the paper is as follows. In section \S\ref{sec:gralmodel}
we describe the general model, which we then calibrate against a fiducial
numerical simulation from \citet{VS+10} in 
\S\ref{sec:calibration}. In \S\ref{sec:obs} we compare the calibrated
model against various observational properties of both large and small
MCs, parameterized only by their mass. In \S\ref{sec:discussion} we
present a discussion, and finally a summary and our conclusions in
\S\ref{sec:sum}.

\section{The Model} \label{sec:gralmodel}

\begin{figure}[!h] 
\begin{centering}
\begin{tabular}{c}
\epsfig{file=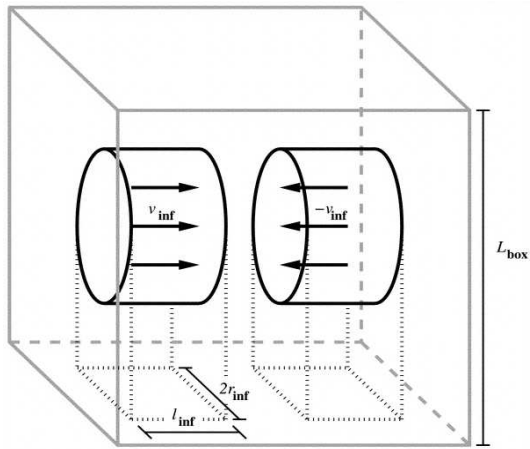,height=6cm,width=7cm}
\end{tabular}
\caption{{\label{fig:fig1-VS07}  Model setup. Cylindrical streams of WNM
are assumed to collide head-on to form first a flattened CNM cloud,
which proceeds to collapse, becoming molecular and star-forming in the
process. The main parameters of the model are the radius of the
cylinders $\Rinf$, and the density and velocity of the inflowing warm gas.}} 
\end{centering}
\end{figure}

We construct a model for studying the SFE in a thin cylindrical cloud
undergoing gravitational contraction, as observed in various simulations
\citep{VS+07, VS+10, VS+11, HH08}. The system is schematically
illustrated in Fig.\ \ref{fig:fig1-VS07}. In the simulations, the collision
of warm neutral medium streams nonlinearly triggers thermal instability,
forming a thin cloud of cold atomic gas
\citep[e.g.,][]{HP99,KI00,KI02,WF00}, which becomes turbulent by the
combined action of various dynamical instabilities \citep{Hunter+86,
Vishniac94, KI02, Heitsch+05, VS+06}. The cloud soon begins to contract
gravitationally as a whole. However, before this global collapse is
completed, some local, nonlinear (i.e., large-amplitude) density
enhancements produced by the initial turbulence manage to collapse on
their own, since their local free-fall time is shorter than the average
one for the entire cloud
\citep{HH08, Pon+11}. These local collapses thus involve only a fraction
of the cloud's total mass. Also, we assume that the newly-formed stars
feed energy back into the cloud. We only consider the ionizing radiation
from massive stars, since this radiation is probably the dominant
mechanism of stellar energy injection at the scale of GMCs
\citep{Matzner02}.

\medskip

In what follows, we investigate the competition between the cloud's
gravitational contraction and its destruction by the mass consumption by
star formation (SF) as well as by the ionization produced by the newly
formed massive stars. Below we describe how we calculate the
contributions from these processes. It is important to note that we do
not follow the chemistry, but rather consider that all the cold gas, either
atomic or molecular, is involved in the gravitational contraction and,
eventually, star formation.

\subsection{Mass accretion} \label{sec:accretion}

In our scenario, the cloud's mass ($\Mc$) at time $t$ is given by
\beq \label{eq:ecdeev}
M_{\rm C}(t)=\int_0^t \dot{M}_{\rm inf}(t')~ {\rm d}t' - M_{\rm
S}(t)-M_{\rm I}(t),
\eeq
where $\dot{M}_{\rm inf}(t)$ is the mass accretion rate onto the cloud
from the WNM inflows, $M_{\rm S}(t)$ is the total mass in stars, and
$M_{\rm I}(t)$ is the mass ionized by stellar feedback. Note that all these
masses are considered to be functions of time. We assume that
\beq
\dot{M}_{\rm inf}(t)=2 \,\rho_{\rm inf} \, v_{\rm inf} \, (\pi \,
R_{\rm C}^2), 
\label{eq:accr_rate}
\eeq
where $v_{\rm inf}$ is the inflow speed, $\rho_{\rm
inf}=n_{\rm inf} \, \mu_{\rm H} \, {\rm m}_{\rm H}$ is the inflow
mass density (with $n_{\inf}$ the number density of the inflows,
$\mu_{\rm H}$ the mean atomic weight of the diffuse gas, and ${\rm
m}_{\rm H}$ the atomic hydrogen mass). Note that $\pi \, R_{\rm C}^2$ is
also the 
cross-sectional area of the thin, cylindrical, cold and dense cloud that
forms by thermal instability at the layer compressed by the inflows. We
take the cloud's radius $R_{\rm C}(t)$ as being initially equal
to the inflow radius $R_{\rm inf}$, and to later decrease as the cloud
contracts gravitationally. The factor of 2 in eq.\ (\ref{eq:accr_rate})
represents the fact 
that there are two inflows, one on each side of the forming cloud.

\subsection{Mass in Stars} \label{subsec:massinstars}

We assume that the SFR is given by the ratio of the gas mass in the
high-density tail ($n > n_{\rm SF}$) of the density distribution
produced by the turbulence in the 
cloud, to its local free-fall time, $t_{\rm ff}(n_{\rm SF})=\sqrt{3
\pi / 32 G \mu m_{\rm H} n_{\rm SF}}$. We refer to $n_{\rm SF}$ as the
threshold density for star formation, and denote by $f(t)$ the fraction
of the cloud's mass that is at densities above $\nsf$ (discussed in
\S\ref{sec:SF_mass_frac}). We then have
\beq \label{eq:SFR}
{\rm SFR}(t)=\frac{M_{\rm C}(t)}{t_{\rm ff}(\nsf)} \, f(t),
\eeq
so that the mass in stars at time $t$ is
\beq \label{eq:Ms}
M_{\rm S}(t)=\int_0^t ~{\rm SFR}(t') {\rm d}t'=\int_0^t ~\frac{M_{\rm
C}(t')}{t_{\rm ff}(\nsf)}~ f(t') ~{\rm d}t'. 
\eeq

We assume that the initial density of the cloud is that of the cold
neutral medium (CNM), in balance with the sum of the thermal and ram
pressures of the inflows, as described in \citet{VS+06}. Typically,
$n_{\rm CNM} \approx 100 \, {\rm cm}^{-3}$, and we use a mean molecular
weight of $\mu=2.35$, adequate for molecular gas. As the cloud evolves
by contraction, the mean density evolves as determined by its mass and
size, so that $\bar \rho = \Mc/ \pi \Rc^2 h$, where $h$ is the cloud
thickness (cf.\ \S \ref{subsec:collapse}).

\subsection{Ionized Mass} \label{subsec:cloudionization}

To model the cloud evaporation by massive stars, we use the results from
\citet{Franco+94}. These authors found that the cloud evaporation rate
by a massive star near the cloud surface is

\beq \label{eq:F94b}
\dot{M}_{\rm I,sur}(t) \approx 2 \pi R_{\rm S,0}^{2} m_{\rm H}
\bar{n} c_{\rm s,I} \, \Big( 1+\frac{5c_{\rm s,I}~t}{2R_{\rm S,0}}
\Big)^{1/5}, 
\eeq
where $t$ is the age of the massive star, $c_{\rm s,I}$ is the sound
speed in the ionized gas, $\bar{n}=\bar{\rho}/\mu m_{\rm H}$ is the
mean number density of the molecular cloud (MC), and
$R_{\rm S,0}$ is the initial Str\"omgren radius of the
massive star in the cloud ($R_{\rm S,0} =[3 S_* /4 \pi \alpha_{\rm B} (2
\bar{n})^2]^{1/3}$, with $S_*$ a representative value of the UV
Lyman-continuum photon flux \citep{Franco+94}, and $\alpha_{\rm B}$ the
recombination coefficient for the ionized gas). We assume that $R_{\rm
S,0}$ is reached immediately at the instantaneous mean density of the
cloud, which, in our model, is continually increasing as the cloud contracts.
Therefore, over a short time interval $\Delta t$ (between $t$ and
$t+\Delta t$), over which the cloud's density can be assumed
constant, a massive star near the cloud's surface can ionize a mass $\Delta
M_{\rm I,sur}= \dot{M}_{\rm I,sur}(t) \Delta t$.

%

In addition, we consider that, during the time interval $\Delta t$, the
cloud forms $\Delta N_{\rm OB}(t) = x_{\rm OB} {\rm SFR}(t) \Delta t /
\langle M_{\rm OB} \rangle$ new massive stars of average mass $\langle
M_{\rm OB} \rangle$, where $x_{\rm OB}$ is the mass fraction of massive
stars, which we calculate assuming an IMF from \citet{Kroupa01}, with
lower and upper mass limits of 0.01 and $60 \, \msol$ respectively.
Defining a star as ``massive'' if it has a mass $M \ge 8 \Msun$, we
obtain $\xob = 0.12$ and a mean massive-star mass $\langle M_{\rm
OB}\rangle = 17 ~\msol$.  Furthermore, in eq.\ (\ref{eq:F94b}) we take
$S_*=2 \times 10^{48} \, {\rm s}^{-1}$, 
$c_{\rm s,I}=12.8 \, {\rm km} \, {\rm s}^{-1}$, and $\alpha_{\rm B}=2.6
\times 10^{-13} \, {\rm cm}^{-3} \, {\rm s}^{-1}$ where we have assumed
that the temperature of ionized gas in the HII region is $10^4
\, {\rm K}$.

With the above considerations, and discretizing the time variable,
the ionized mass at time $t_i$ is given by
\beq
\MI(t_i) = \MI(t_{i-1}) + \sum_{j=k}^i \dnob(t_j) \Delta M_{\rm I,sur}(t_j),
\label{eq:ion_mass}
\eeq
where $t_k$ is the time at which the oldest remaining OB stars were
formed, $k = i - {\rm int} \left(t_{\rm OB}/\Delta t\right)$, `int' is the
integer function, and $\tob = 5$ Myr is the main-sequence lifetime of our
representative OB star. The second term
in the right-hand side of eq.\ (\ref{eq:ion_mass}) thus gives the mass
ionized over the time interval $\Delta t = t_i -t_{i-1}$ by the stars
formed between $t_k$ and the present time, $t_i$. Note that we have
neglected the possibility that part of the ionized gas can recombine and
return to the cloud, as we expect the ionized gas to escape to the
diffuse medium.

Finally, to account for the fact that the massive stars are born
in dense environments, in eq.\ (\ref{eq:F94b})
we take the density as the maximum between the instantaneous cloud
density $n(t)$  and $10^3 \pcc$
(typical of clumps), in order to avoid an over-ionization when the cloud
density is low.

\subsection{Global gravitational collapse.} \label{subsec:collapse}

Following the trend seen in the numerical simulations, we assume that
our clouds evolve in two stages. First, a mass-growth stage occurs,
during which the cloud (initially of zero mass) increases its mass by
accretion from the WNM at constant radius and density, so that only its
thickness increases \citep[see, e.g.,][]{VS+06, VS+07, FW06}, until it
reaches its thermal Jeans mass. At that point, the second stage begins,
during which the cloud undergoes global gravitational contraction. For
circular modes in a self-gravitating isothermal sheet of finite
thickness, the Jeans mass is given by \citep{Larson85}:
\beq \label{eq:MJ}
M_{\rm J}=4.67 \frac{c_{\rm s}^4}{G^2 \Sigma}
\eeq
where $c_{\rm s}$ is the cloud sound speed, and $\Sigma = M(t) / \pi
R^2(t)$ is the surface density, with $M(t)$ and $R(t)$ being the
instantaneous mass and radius of the cloud, respectively. In order to
calculate the sound speed we first compute the cloud's temperature. To
do this, we use the fit by Koyama \& Inutsuka (2002; see also the note
in \VS\ et al. 2007) to the heating and cooling processes considered by
\citet{KI00}. This allows us to solve for the temperature of thermal
equilibrium (heating = cooling) as a function of the density. In this
way, we get temperatures of $T \approx 42 ~ {\rm K}$ for a density of $n=
10^2 ~ \pcc$ and $T \approx 7 ~ {\rm K}$ for $n \sim 10^7 ~ \pcc$.

During the mass-growth stage, over which the cloud's
radius remains constant ($R = R_{\rm inf}$), the
cloud's thickness $h$ is given by the condition of constant number density:
\beq
h(t)=\frac{M_{\rm C}(t)}{100 \, \mu \, m_{\rm H} (\pi \, R_{\rm inf}^2)}.
\eeq
Once the collapse begins, we assume that the thickness remains constant
at the final value achieved during the growth stage, and that its volume
density increases only due to the radial contraction, as suggested by
simulations including self-gravity \citep[e.g.,][]{VS+07, VS+10, VS+11,
HH08}. This assumption is equivalent to assuming that the average
thickness of the cloud is much smaller than its Jeans length throughout
its evolution. In general, this is a good approximation until when the
cloud has contracted to radii of a few pc.

To determine the radial evolution during the contraction stage, we first
calculate the acceleration at the cloud's edge. We take a reference frame with
its origin at the cloud's center and with its $x$-axis along the inflow
direction (i.e., perpendicular to the plane of our flattened cloud). We
integrate over mass elements $\bar{\rho}~dz dy dx$ located at a distance
$[(R-z)^2+y^2+x^2]^{1/2}$ from the edge. At 
a certain time $t_i$, the acceleration at the cloud edge is
\beq
a(t_i)=2 \, G \, \bar{\rho}_i \, \int_{-R_i}^{R_i} \, dz \,
\int_{0}^{\sqrt{R_i^2-z^2}} dy \, \int_{-h/2}^{h/2} \,
\frac{R_i-z}{[(R_i-z)^2+y^2+x^2]^{3/2}} \, dx 
\eeq
where $\bar{\rho}_i=\bar{\rho}(t_i)$ and $R_i=R_{\rm C}(t_i)$. We solve
the first integral analytically, while the second and third ones are solved
numerically by the composite Simpson rule. After a small time increment
$\Delta t=t_{i}-t_{i-1}$, the change in the cloud radius is:
\beq
R_{i+1}=R_i-v_{0,i} \,\Delta t-\frac{1}{2} a(R_i)  \Delta t^2
\label{eq:R(t)}
\eeq
where the instantaneous velocity at time $t_i$ is
$v_{0,i}=\sum_{j=0}^{i} a(R_j)\Delta t$ for constant $\Delta t$.

Finally, we introduce a correction factor, representative of the fact
that true gravitational collapse of a gaseous mass does not occur in
strict free-fall, since thermal pressure is never completely negligible,
as pointed out in the pioneering work by \citet[][ Appendix
C]{Larson69}. There, he reported that the collapse of his simulations
occurred in a time longer than the free-fall time by a factor of
1.58. Thus, at each timestep, we divide the radius given by eq.\
(\ref{eq:R(t)}) by a factor $\fL$, which we refer to as the ``Larson
parameter'', and calibrate in \S \ref{sec:calibration}.

\subsection{Density distribution}
\label{subsec:PDF}

A key ingredient in our evolutionary model is the fraction of dense gas
that is participating in the SF process, and therefore so is the
evolution of the {\it probability density function} (PDF) of the cloud's
density field.

In our model, we consider that our clouds are born transonically
turbulent, as a consequence of the various instabilities at play in the
compressed layer between the streams \citep{Vishniac94, WF00,
Heitsch+05, Heitsch+06, VS+06}. Also, in the absence of direct stellar
irradiation, the temperature in the cold gas
varies at most by factors of a few for densities $100 < n < 10^7 \pcc$,
and thus, as a first approximation, we consider it to be isothermal.
Therefore, we assume that the density field within the cloud
is initially characterized by a lognormal PDF,
appropriate for supersonically turbulent, isothermal gas
\citep{Passot-VS98}. The PDF is then given by
\beq \label{eq:pdf}
P(s)=\frac{1}{\sqrt{2\pi \sigma_s^2}} ~ {\rm exp} \left[- \frac{(s -
s_{\rm p})^2}{2\sigma_s^2} \right], 
\eeq
where 
\beq
s \equiv \ln (\rho / \bar \rho),~~~ s_{\rm p}= \ln (\rho_{\rm p} /\bar
\rho)=-\sigma_s^2/2,
\label{eq:mean_PDF}
\eeq
with $\rho_{\rm p}$ the peak density, and 
\beq
\sigma_s^2= \ln (1+ b^2 \mathcal{M}_{\rm rms}^2),
\label{eq:stddev_PDF}
\eeq
where $b$ is a proportionality constant related with the compressibility
induced by the turbulent forcing, which for simplicity we take equal to
unity \citep[see e.g.,][]{VS94, Padoan+97, Passot-VS98, Federrath+08}.

However, more recent numerical and observational studies suggest that
the density PDF in gravitationally contracting systems does not preserve
its lognormal shape during the collapse, but rather develops a power-law
tail at high densities \citep{Klessen00, DB05, VS+08, Kainulainen+09,
Kritsuk+11, BP+11b}. In particular, \citet{Kritsuk+11} have suggested
that the final slope should be in the range [3/2,7/4], but at the
present time we know of no theoretical prediction as to how the slope
nor the transition point between the lognormal and the power law should
evolve in time. We experimented with various options for modeling the
evolution of the PDF's power-law tail, but found the behavior to be very
sensitive to the parameters used, while we had no physical ways of
constraining them.  Finally, \citet{Kritsuk+11} proposed that the origin
of the power-law tail is the development of local collapsing flows, that
may have either Larson-Penston \citep{Larson69, Penston69} or
\citet{Shu77} density profiles. In this case, the power-law tail is an
{\it effect} of the gravitational collapse rather than its {\it cause},
and thus it should not be counted as providing turbulent seeds for
future collapses. For all of these reasons, we do not consider the power-law
form of the PDF, and stick to the lognormal.

It is worth noting that the density PDF we consider here refers
only to the cold (approximately isothermal) gas that makes up the cloud,
and not to the entire gas contents of the system. This means that this
PDF is not directly comparable to that observed in the numerical
simulations of the same process, which corresponds to thermally-bistable
gas. Thus, we did not consider the possibility of using a PDF extracted
from numerical simulations, either.

We model the evolution of the {\bf lognormal} density PDF as follows. First, as
indicated by eq.\ (\ref{eq:mean_PDF}), the mean of the PDF is given by
the cloud's mean density, given by $\bar \rho = \Mc/ \pi \Rc^2 h$. In
turn, we prescribe that $\bar \rho$ varies as follows. During the
mass-growth stage, it remains constant, at $100 \pcc$. Once the
contraction stage begins, it increases, causing the
density PDF to shift to higher values. Eventually, however, the SFR
becomes large enough that the ionization by newly born massive stars
reduces the cloud's mass rapidly enough as to cause the PDF to shift
back towards lower densities again.

Second, the standard deviation of the PDF is determined by the turbulent
rms Mach number, as indicated by eq.\ (\ref{eq:stddev_PDF}).
Unfortunately, the evolution of the {\it turbulent} component of the
Mach number remains rather uncertain. Standard relations, such as
Larson's (1981) velocity dispersion size cannot be assumed here. Indeed,
in our collapsing-cloud scenario, the majority of the velocity
dispersion is due to the contracting motions \citep{BP+11} rather than
to random turbulent motions, and should not be counted as turbulence
capable of producing new density fluctuations suceptible of subsequent
collapse.  

Of course, it is natural to assume that a fraction of the kinetic energy
in the collapsing motions will be transferred to random motions, but
this problem remains largely unexplored in the case of gaseous
media. \citet{VS+98} numerically investigated the scaling of the
non-collapsing component of the velocity dispersion in collapsing
spherical clouds. Those authors found that the turbulent velocity
dispersion scaled as $\rho^x$, with $x \in [1/4,1/2]$, depending on the
particular setup of the collapse and on the presence of magnetic
fields. However, that study was performed at low resolution and was
restricted to spherical geometry, so it cannot be taken as definitive.
More recently, \citet{KH10} suggested that the rate of kinetic energy
injection by the warm neutral streams feeding a molecular cloud is at
least one order of magnitude larger than the rate of turbulent energy
dissipation within the clouds. However, these estimates are not enough
to properly constrain the evolution of the turbulent kinetic energy in
our model clouds. Thus, we simply take a {\it constant} value of the
turbulent Mach number, which represents a compromise between turbulent
decay by dissipation and feeding of the turbulence by transfer from the
collapsing motions. This thus implies that the width of the PDF in our
model remains constant through the cloud's evolution as well. We take
this constant value of the turbulent Mach number as $\mathcal{M}_{\rm
rms}=3$, the canonical value for the cold neutral medium \citep{HT03}.

\subsection{Star-forming mass fraction} \label{sec:SF_mass_frac}

Finally, as mentioned in \S \ref{subsec:massinstars}, we assume that
only gas with number density higher than a threshold value $n_{\rm SF}$
participates in the {\it instantaneous} SF process. This amounts to assuming
that $\nsf$ is sufficiently larger than the cloud's mean density $\bar
n$ as to guarantee that $\tff(\nsf) \ll \tff(\bar n)$, where $\tff(n)$
is the free-fall time corresponding to density $n$. In a sense, this 
prescription may be thought of as the model's analogue of sink particles
in a numerical simulation \citep{Bate+95, Federrath+10}, in which gas at
sufficiently high densities is replaced by point mass particles
representing collapsed objects.

Thus, for the assumed lognormal PDF the fraction of dense gas that forms
stars is given by
\beq \label{eq:f}
f(t)=
\frac{1}{2} \left[ 1-{\rm erf} \left( \frac{2 \,
s_{\rm SF}(t)-\sigma_s^2}{\sqrt{2} \sigma_s} \right) \right],
\eeq
where $s_{\rm SF}={\rm ln}(\rho_{\rm SF}/\bar \rho)$, and $\rho_{\rm
SF}$ is the volume density corresponding to $\nsf$ \citep[see
also][]{Elmegreen02, KM05, Dib+11}. {\bf The value of $\nsf$ is
calibrated against a numerical simulation by \citet{VS+10} in \S
\ref{sec:calibration}.}

\subsection{Temporal evolution} \label{sec:time_step}

With all the above ingredients, we discretize eq.\
(\ref{eq:ecdeev}) as
\beq \label{eq:ecdeev2}
M_{\rm C}(t_i)= \sum_{j=0}^i \dot{M}_{\rm inf}(t_j)\Delta t -
\sum_{j=0}^i  \frac{M_{\rm C}(t_j) \, \Delta t}{t_{\rm ff}(t_j)} \,
f(t_j) - \Big[ M_{\rm I}(t_{i-1})+\sum_{j=k}^i \Delta N_{\rm OB}(t_j)
\Delta M_{\rm I,sur}(t_j) \Big] 
\eeq
We integrate this equation numerically over time, taking $\Delta t=4 \times
10^{-4} \, {\rm Myr}$. We choose this value as a reasonable compromise
between speed and accuracy, after experimenting with timesteps down to
0.01 times this value, from which we found (with arbitrary free parameters) 
that the chosen value gives already well-converged values of the final cloud 
masses, the variable that turned out to be most sensitive to the size of the 
timesteps.

We consider that the cloud's evolution ends either because the cloud
gets completely evaporated by the massive-star ionization or because its
average density reaches $\nsf$, a point at which the remaining mass is
fully converted into stars and the cloud disappears.

\section{Calibration of the model} \label{sec:calibration}

With all the model ingredients defined, we now proceed to calibrate it
by matching it to one of the numerical simulations by \citet{VS+10}. We
consider the simulation labeled SAF1 in that paper, which most resembles
the physical conditions built into our model. The label SAF indicated
that it contained small-amplitude initial velocity fluctuations ($\sim
2$\% of $\vinf$), which were sufficient to trigger the instabilities
responsible for turbulence production in the forming cloud, but not
large enough to significantly distort its sheet-like geometry. 

\begin{figure}[!ht] 
\begin{tabular}{c}
\plottwo{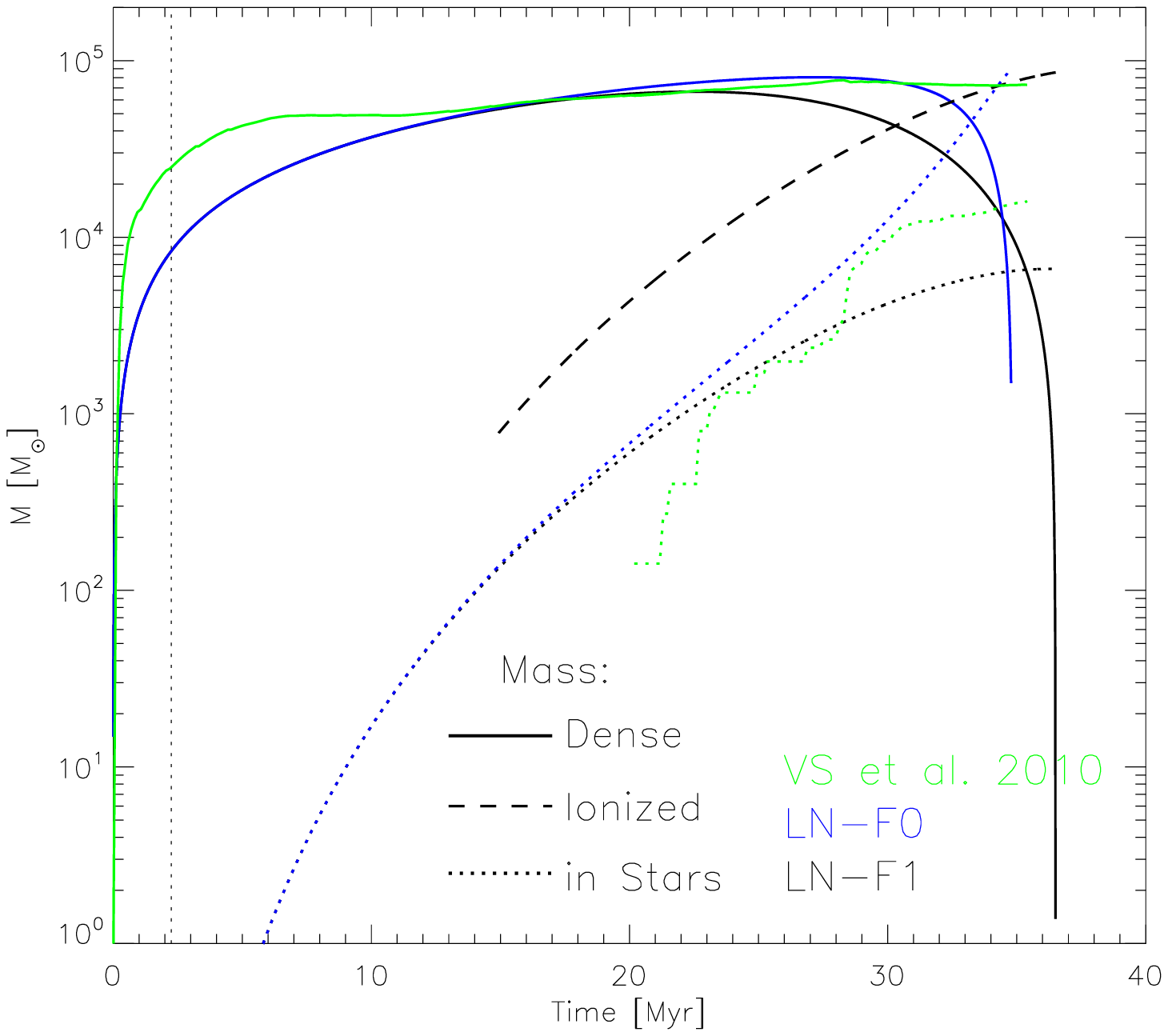}{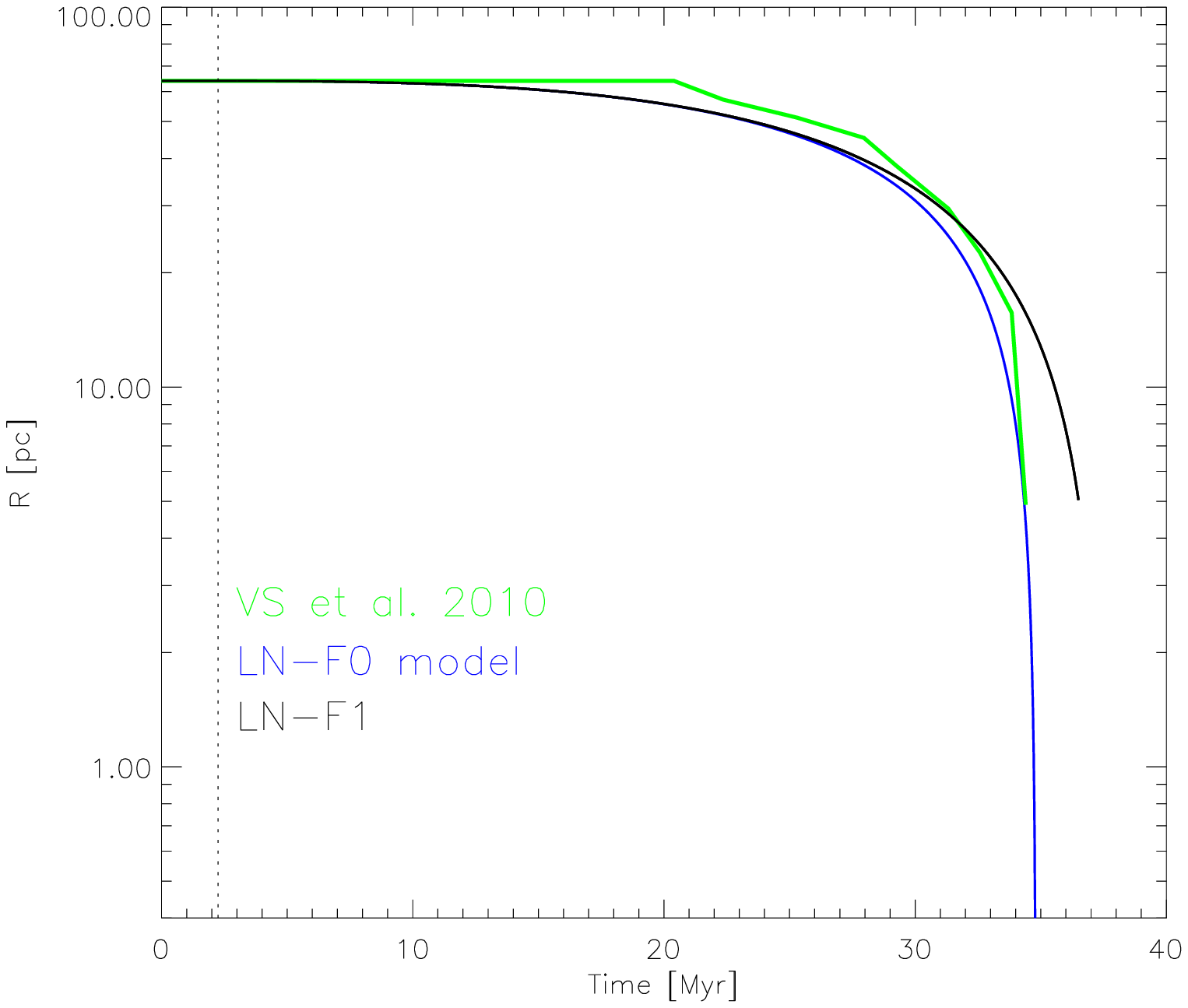}
\end{tabular}
\caption{{\it Left:} Evolution of the dense gas mass, ionized gas mass, and
mass in stars for the model cloud with parameters that best match the
corresponding quantities (except for the ionized mass, which is not
measured) in the SAF1 simulation of \citet{VS+10}. The green lines
represent the simulation, the blue lines represent the model cloud with
stellar feedback turned off, and the black lines represent the model
cloud with the stellar feedback turned on, with $\nsf = 10^6 \pcc$. The
vertical line represents the beginning of global collapse. {\it Right:}
Evolution of the cloud radius in the model and in the simulation, with
the same color coding as in the left panel.}
\label{fig:mass_radius}
\end{figure}

The simulation considered the collision of two cylindrical WNM streams
with $n_{\rm inf}=1 \, {\rm cm}^{-3}$, $T_{\rm inf}=5000 \, {\rm K}$,
inflow radius of $R_{\rm inf}=64 \, {\rm pc}$, and an initial inflow
speed $v_{\rm inf}=7.5 \kms$.\footnote{Note that there are typographical
errors in the numbers reported by \citet{VS+10}. The values given here
are the correct ones.} Thus, for our calibration purposes, we use these
same values for our model, except for $\vinf$. We do this because, in
the simulation, the inflow speed decreases over time, because the rear
end of the inflowing cylinder leaves a vacuum behind it. As a
consequence, in the reference frame of the inflow, the inflow's rear end
tends to re-expand into this vacuum, producing a velocity gradient along
the cylinder. Instead, in the model we use a constant inflow speed, and
thus we take a smaller value, representative of the time-averaged speed
in the simulation. We find that a value of $\vinf = 4.5 \kms$ provides
the best match for the evolution of the simulation cloud's mass. Also,
we find that the temporal evolution of the cloud's radius is best
matched with a value of the Larson parameter of $\fL = 1.7$, very
close to the original value of 1.58 originally proposed by
\citet{Larson69}. We show the evolution of these quantities for both the
simulation and the model in Fig.\ \ref{fig:mass_radius}. Note that the
evolution for the model with the ionizing stellar feedback turned both
on and off is shown. We refer to the model with feedback off as the
LN-F0 case, and to the one with the feedback on, as the LN-F1 case.

\begin{figure}[!ht] 
\begin{tabular}{c}
\plottwo{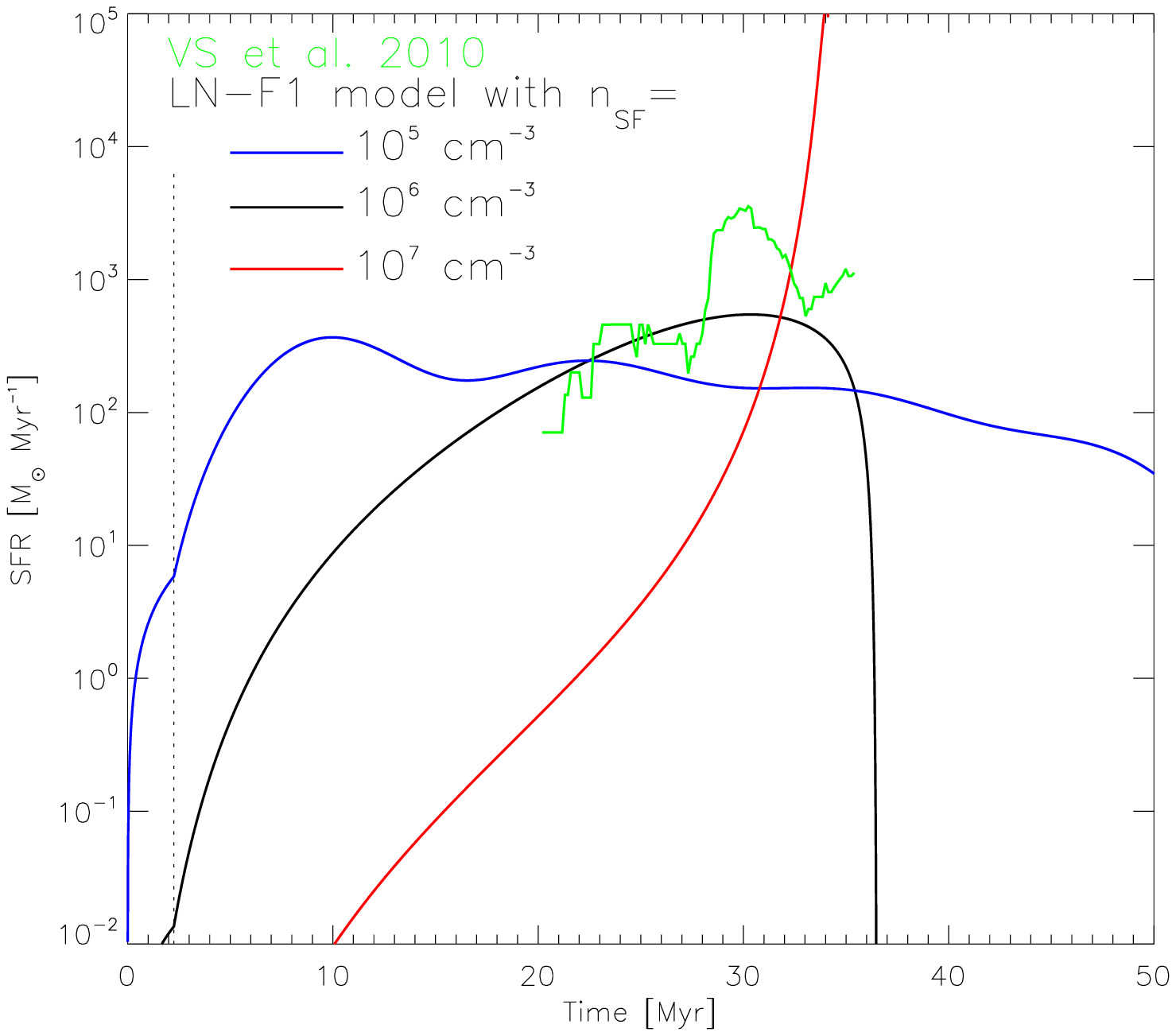}{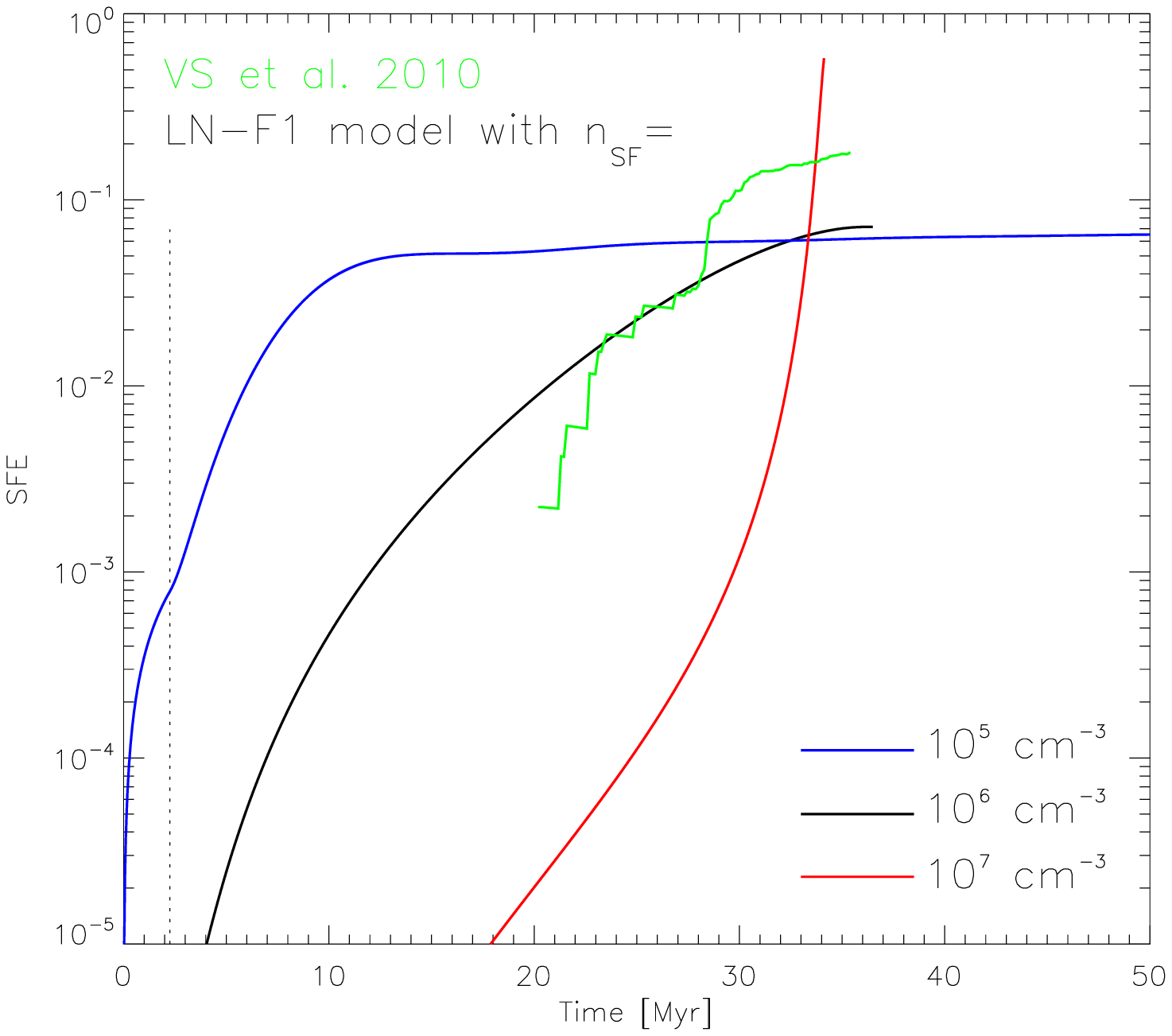}
\end{tabular}
\caption{{\it Left:} Evolution of the star formation rate (SFR) for the
model cloud for the cases $\nsf = 10^5$, $10^6$, and $10^7 \pcc$,
compared to the evolution of the SFR in the SAF1 simulation of
\citet{VS+10}. {\it Right:} Same as the left panel, but for the star
formation efficiency (SFE). The green lines represent the numerical
simulation.}
\label{fig:SFR_SFE}
\end{figure}

Next, keeping the above parameters fixed, we vary the $\nsf$ parameter,
the density threshold for star formation, to match the SFR and SFE of
the simulation, with the SFE being defined as
\beq
{\rm SFE(t)} = \frac{\Ms(t)}{\Mc(t) + \Ms(t) + \MI(t)},
\label{eq:SFE}
\eeq
where $\Mc$ is the dense gas mass, $\Ms$ is the total mass in stars,
$\MI$ is the total mass ionized by stars, all quantities being
time-dependent. Figure \ref{fig:SFR_SFE} shows the evolution of the SFR
and SFE of the model for $\nsf = 10^5$, $10^6$, and $10^7 \pcc$, and
compares them with the evolution of the corresponding quantities in the
simulation, showing that the best match is obtained with a value $\nsf =
10^6 \pcc$, which we use in the rest of the paper. This value is
reassuring since, on the one hand, it is comparable to the
sink-formation density threshold used in the numerical simulations
\citep{VS+07, VS+10}, and on the other, it is high enough that the
material above those densities can be safely assumed to be locally
gravitationally bound \citep{Galvan+07, HH08}.

\section{Discussion of the calibrated model's evolution} 
\label{sec:calibr_evol}

Once the model parameters have been calibrated with the SAF1 simulation,
it is illustrative to discuss the general features of the model cloud's
evolution, which hold qualitatively for the other cases we explore in \S
\ref{sec:obs}, where varying only the inflow radius $\Rinf$ we are able
to match and explain several observed features of molecular clouds and
their star-forming activity.

First, we note from the {\it right} panel of Fig.\ \ref{fig:mass_radius}
that the radius of the model with feedback (LN-F1) evolves slightly more
slowly than that of the case without feedback (LN-F0). This is because
the stellar feedback erodes the cloud through ionization, thus reducing
its mass, which in turns causes a lower gravitational acceleration, thus
slowing the collapse. We also note that the radius of the LN-F0 case
approaches zero at late times, while that of the LN-F1 ends at a finite
radius, implying that the evolution is terminated because the cloud is
completely evaporated before it reaches zero radius (see below). 

Second, from the {\it left} panel of Fig.\ \ref{fig:mass_radius}, we
note that the dense gas mass of {\it both} the LN-F0 and LN-F1 models
decreases at late times. This is because even in the non-feedback LN-F0
model, gas is consumed due to the conversion of gas to stars. However,
we see that the mass consumption is much more abrupt in the LN-F0 model,
while in the LN-F1 model it proceeds more slowly. This indicates that
the feedback inhibits star formation to the level that the mass consumed
by ionization from the feedback is {\it less} than the mass that would
be consumed by star formation were there no feedback. Nevertheless, note
that the dense gas mass in the LN-F1 model approaches zero at the end of
the evolution, indicating that all of the cloud's mass is used up by the
combined action of ionization and star formation. This corresponds to
the well known fact that clusters eventually destroy their natal cloud,
and are left with no gas around them after several million years
\citep{Leisawitz+89}.

\begin{figure}[!ht] 
\plottwo{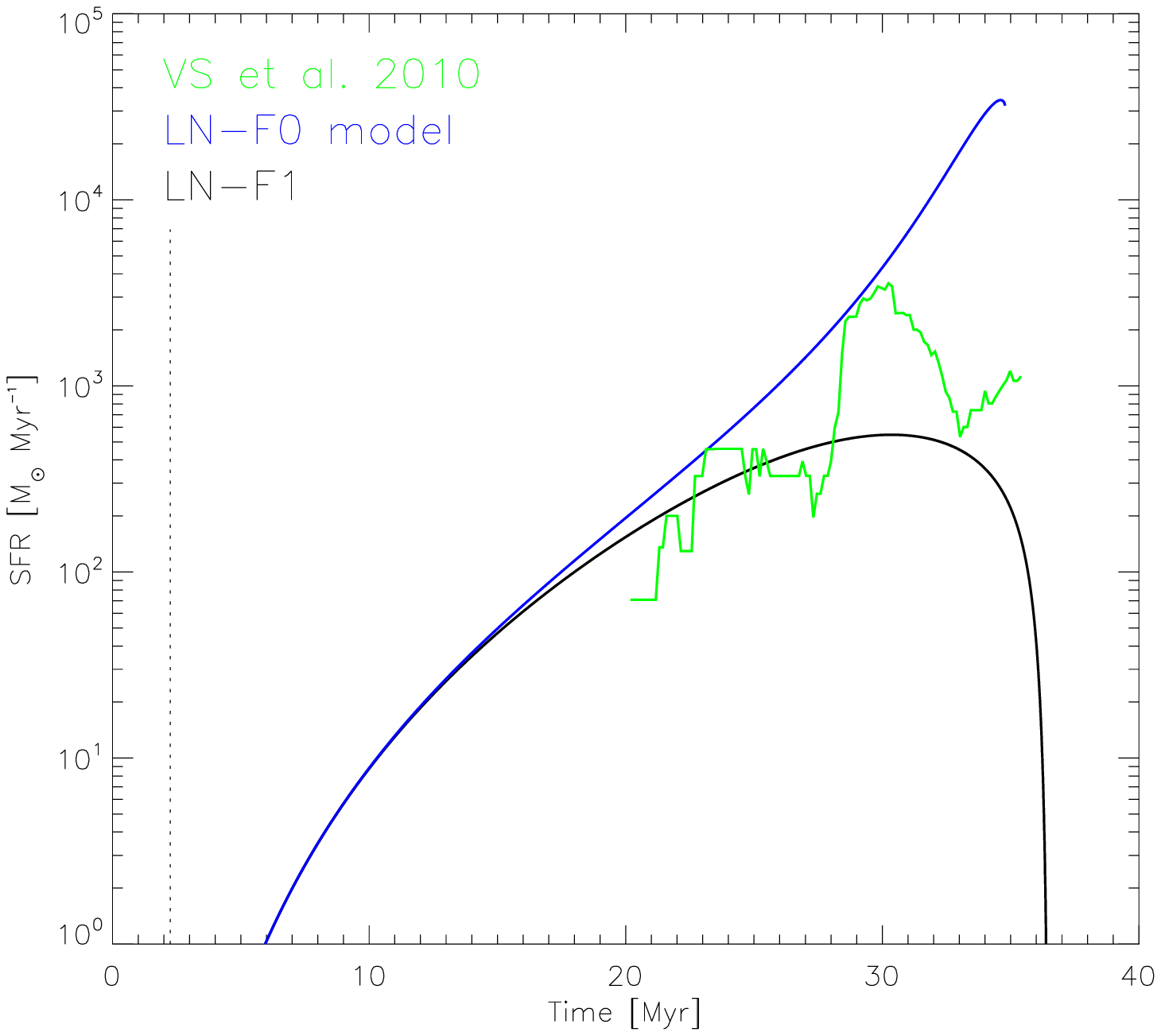}{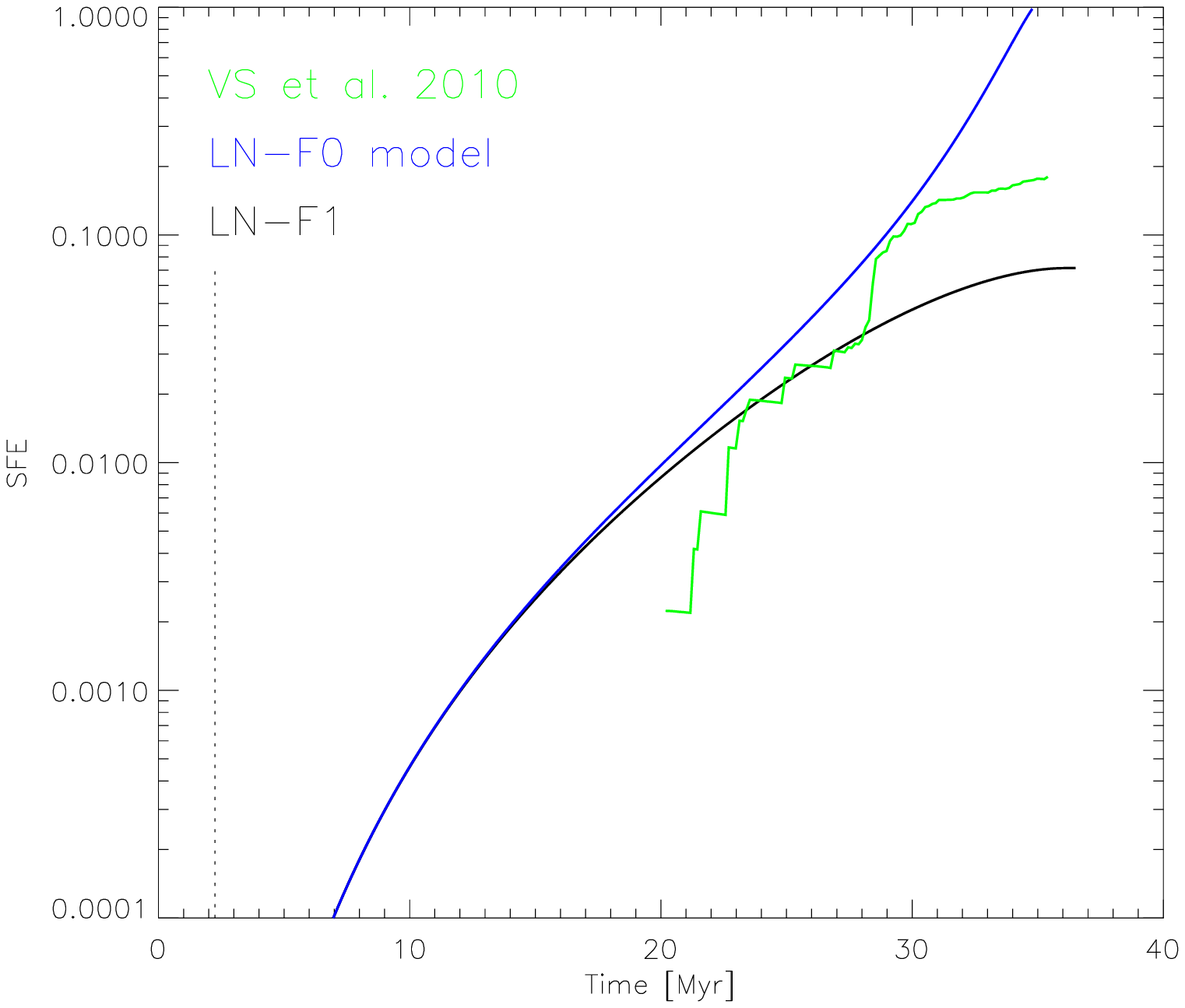}
\caption{Evolution of the SFR ({\it left}) and the SFE ({\it right}) of
the calibrated model with $\nsf = 10^6 \pcc$ for a case with feedback
turned off (LN-F0, {\it blue line}) and a case with feedback on (LN-F1,
{\it black line}). The green lines represent the numerical
simulation. The LN-F0 model has accelerating SFR and SFE, while they
decelerate for model LN-F1. See also discussion in \S \ref{sec:accel_SF}}
\label{fig:SFR_SFE_ne6}
\end{figure}

Finally, note from Fig.\ \ref{fig:SFR_SFE_ne6} that both the SFR and the SFE
increase at an ever faster pace until the end of the simulation in the
non-feedback case LN-F0, while their growth slows down in the case of
LN-F1. In this case, the SFR eventually begins to decrease and goes to
zero at the end of the evolution, as the dense gas mass is completely
consumed by the ionization, leaving no further fuel for star formation.
Instead, the SFR in the LN-F0 case would reach a singularity at a
finite time were it not for the time discretization of our model.

\section{Comparison with observations and previous work} \label{sec:obs}

We now proceed to compare the results of our model with related
observational results. For the various comparisons, we vary only the
inflow radius parameter, $\Rinf$, which determines the maximum dense gas
mass attained by the model. 

\subsection{Evolutionary stages} \label{sec:evol_stages}

As a first case in point, we consider the evolutionary stages of the
clouds. Recently, \citet{Kawamura+09} have suggested that GMCs in the
Large Magellanic Cloud undergo four evolutionary stages. In the first
stage (Type I, with a duration of $\sim 7$ Myr and a median mass $\MtI
\sim 10^{4.8}~\msol$), the GMCs show no sign of massive star
formation. In the second stage (Type II; 14 Myr; $\MtII \sim
10^{5.2}~\msol$), the GMCs have only HII regions, while in the third
(Type III; 6 Myr; $\MtIII \sim 10^{5.4}~\msol$ ), the GMCs contain both
HII regions and clusters.  Finally, the last stage (IV) corresponds to
the time when the GMCs have been completely dispersed, and only young
clusters and/or SNRs are found.

\begin{figure}[!ht] 
\begin{centering}
\epsfig{file=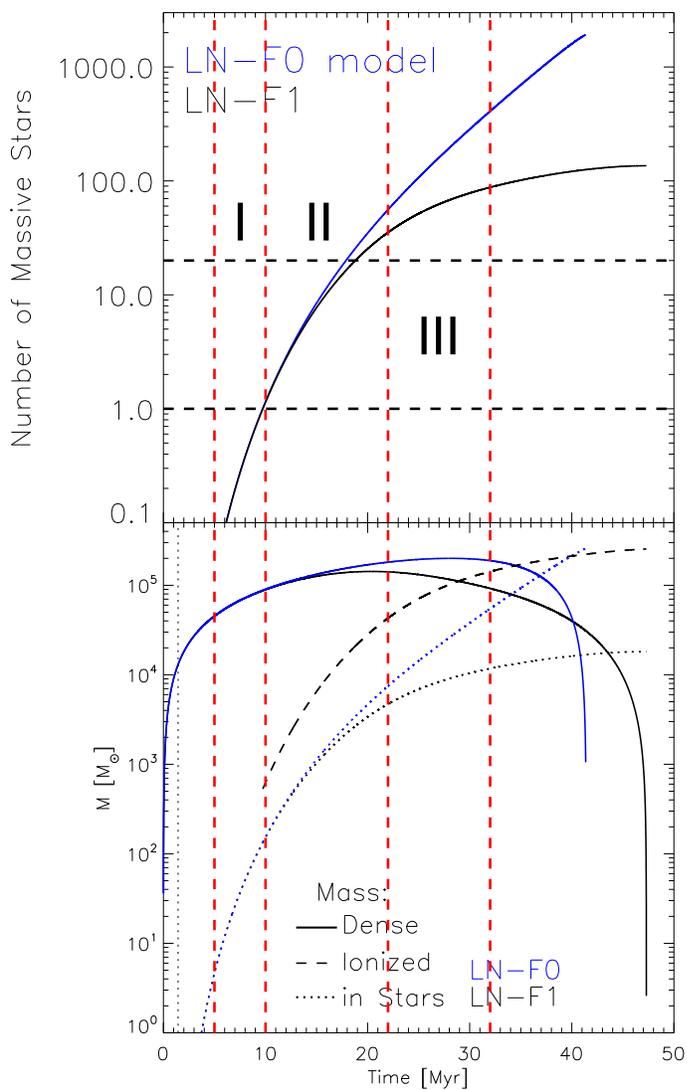,height=15cm}
\caption{{\it Top:} Evolution of the number of massive stars in the
model cloud with $\Rinf = 100$ pc, showing the periods which roughly
correspond to the cloud types defined by \citet{Kawamura+09}.  {\it
Bottom:} Evolution of the dense gas, ionized and stellar masses for this
cloud.}
\label{fig:comp_Kawa}
\end{centering}
\end{figure}

Because the masses of the clouds in their sample are near $10^5
\Msun$, we choose an inflow radius $\Rinf = 100$ pc, which gives a
maximum cloud mass ($\Mmax$) of slightly over $10^5~\Msun$, thus making
it directly comparable to their cloud sample. Figure \ref{fig:comp_Kawa}
shows the evolution of both the number of massive stars in the model
cloud ({\it top panel}) and the masses of the dense gas, ionized gas,
and stellar components of the cloud ({\it bottom panel}). For
comparison with the cloud types defined by \citet{Kawamura+09}, here we
define Type I as the epoch when the model cloud has less than one
massive star, Type II as the period when the cloud has less than 20
massive stars, and Type III as the period when the cloud has more than
20 massive stars. We see that the model cloud spends $\sim 5$ Myr as a
Type I, $\sim 12$ Myr as a Type II, and $\sim 10$ Myr as a Type III,
noting that after such a time the cloud's mass has decreased by more
than a factor of 2, and may be considered to be on its way to
disappearing.

We also note from the {\it bottom panel} of Fig.\
\ref{fig:comp_Kawa}, that the dense gas mass of the model cloud varies
only moderately during the time it spends as an either Type I, II or III
cloud, although a net increase from Type I to Type II is apparent.
Moreover, because the cloud's mass decreases by over a factor of 2 while
it is in the Type III stage, a significant scatter in cloud masses is
expected in this class, as is indeed observed in Fig.\ 12 of
\citet{Kawamura+09}. 

From the above discussion, we thus conclude that the evolution of our
model GMC, with $\Rinf = 100$ pc, compares well, both qualitatively and
quantitatively, with the evolutionary scheme proposed by
\citet{Kawamura+09} for GMCs in the LMC.

\subsection{Fiducial model vs. OMC-1} \label{subsec:OMC1}

Next, we compare our model cloud with the physical conditions of real
star-forming regions. As an example, we
choose the clump known as OMC-1 in the Orion Molecular Cloud, which can
be considered a typical massive star-forming region, and has been
extensively studied. The total gas mass in OMC-1 is $\sim 2200\, \msol$
\citep{Bally+87}, the size is $\sim 1.35 \, {\rm pc}$ (which implies a
number density $\sim 1.54 \times 10^4 \, {\rm cm}^{-3}$), and the mass
in stars is $\sim 500 \, \msol$, which implies that the average SFR is
$\langle {\rm SFR} \rangle \gtrsim 2.5 \times 10^{-4} \, \msol {\rm
yr}^{-1}$, assuming a stellar age spread of $\lesssim 2 \, {\rm Myr}$
\citep[see][and references therein]{VS+09}.

\begin{figure}[!ht] 
\begin{centering}
\begin{tabular}{c}
\epsfig{file=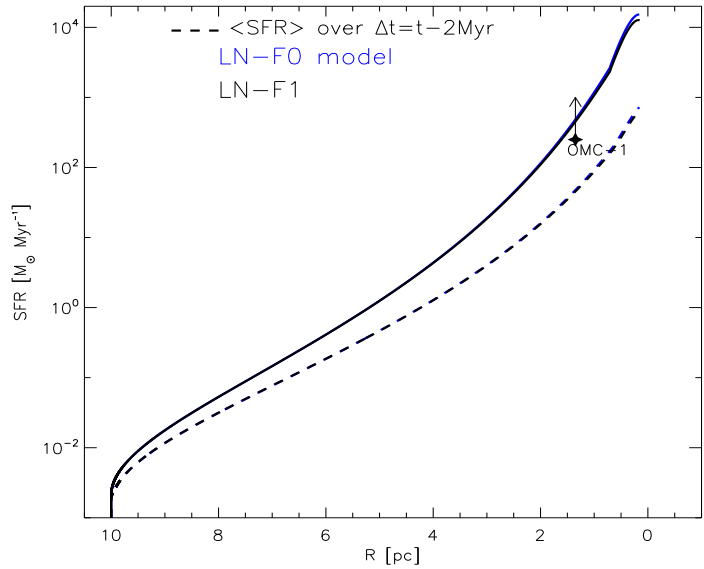,height=8cm,width=10cm}
\end{tabular}
\caption{{\label{fig:SFR-R} } Evolution of a model cloud with $\Rinf =
10$ pc in a SFR vs.\ size ($R$) diagram. Note that the size axis is
reversed, so that the cloud evolves from left to right as it contracts
gravitationally.  Also plotted is the locus of the OMC-1 star-forming
region in this diagram, showing that the model comes very close to that
locus towards the end of its evolution.}
\end{centering}
\end{figure}

To compare with this, we choose a value of $\Rinf = 10$ pc for our
model, which at $t = 27.6$ Myr has the same density as OMC-1, making it
directly comparable to the latter. At this time, the model cloud has a
mass $M \approx 1800~\Msun$, a mass in stars $\approx 200 \, \msol$, and
size (diameter) $\approx 1.9$ pc. Moreover, the average SFR over
the last 2 Myr (the age dispersion used to compute the SFR of OMC-1) is
$\langle {\rm SFR} \rangle \approx 100~\Msun~{\rm Myr}^{-1}$ (see Figure
\ref{fig:SFR-R}). These values are in very good agreement with the
estimates for OMC-1, suggesting that our evolutionary model correctly
describes this cloud.

\subsection{Kennicutt-Schmidt relation.} \label{subsec:kennicutt}

Another possible point of comparison of our model with observational
data is provided by the so-called Kennicutt-Schmidt relation. Ever since
the seminal paper by \citet{Schmidt59}, it has been well known that
there exists a relationship between the SFR and the gas density in
galaxies. Four decades later, collecting data from various surveys of
nearby normal and starburst galaxies, \citet{K98} found a clear
correlation between the galaxy-averaged SFR surface density
($\Sigma_{\rm SFR}$) and the galaxy-averaged total gas surface density
($\Sigma_{\rm gas}=\Sigma_{\rm HI}+\Sigma_{\rm H_2}$, where $\SHI$ and
$\SHtwo$ are the HI and H$_2$ surface densities, respectively) of the
form $\Sigma_{\rm SFR} \propto \Sigma_{\rm gas}^N$, with $N \approx
1.4$.

Recent observations of external galaxies with high spatial resolution
(on scales of fractions of kpc) show that $\Sigma_{\rm SFR}$ scales
almost linearly with $\Sigma_{\rm H_2}$, while no clear correlation
exists with $\Sigma_{\rm HI}$ \citep[see e.g.,] [] {Wu+05,Bigiel+08}.
However, observations of individual clouds \citep[e.g.,] [] {Evans+09,
Heiderman+10} systematically show larger values of $\SSFR$ than those
implied by the fits by \citet{K98}, \citet{Bigiel+08}, and
\citet{Wu+05}. Moreover, the SFRs derived by \citet{Heiderman+10} for
their massive clump sample were obtained using extragalactic methods
\citep[taken from][]{Wu+10}, and they warn that this could cause the SFRs
they report to be underestimated by up to 0.5--1 orders of magnitude,
implying an even stronger disagreement with the galaxy-scale
measurements.

\begin{figure}[!ht] 
\begin{centering}
\epsfig{file=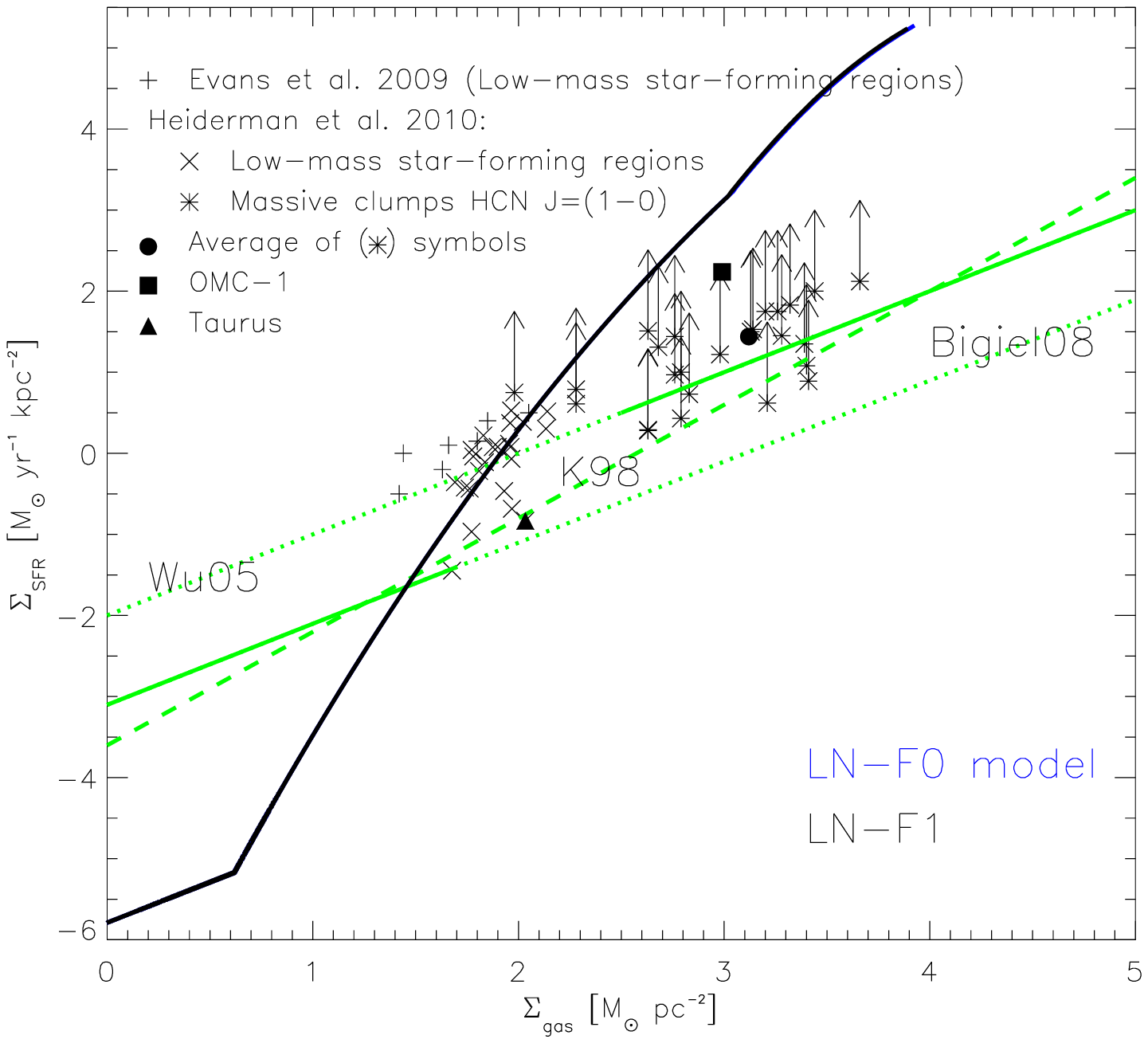,height=8cm,width=10cm}
\caption{{\label{fig:kennicutt} SFR surface density $\Sigma_{\rm SFR}$
vs. gas surface density $\Sigma_{\rm gas}$. The dashed line
represents the {\em Kennicutt-Schmidt relation}, while the lower dotted
line represents the observational fit by \citet{Bigiel+08} and the top
dotted line is the fit by \citet{Wu+05}. We also plot the data for
individual low- to intermediate-mass star-forming regions by
\citet{Evans+09} (+ symbols) and \citet{Heiderman+10} ($\times$ symbols)
and for massive clumps by \citet{Heiderman+10} ($\ast$ symbols). The
filled square represents OMC-1 \citep[see][]{VS+09} and the filled
triangle is Taurus \citep[see, e.g.,][]{Heiderman+10}. The arrows in the
massive clumps from the latter authors indicate the likely correction to
the SFR due to their application of extragalactic methods to Galactic
regions. The solid black line shows the evolution of our calibrated
model with $\Rinf = 10$ pc.}}
\end{centering}
\end{figure}

These cloud-scale observations occupy a well-defined locus in
$\Sigma_{\rm gas}-\Sigma_{\rm SFR}$ space, which can be compared with
our model. For this task, we choose an inflow radius $\Rinf = 10$ pc,
for which our model reaches a maximum mass of $\Mmax \approx 2000
~\Msun$, almost identical to the median mass of the \citet{Evans+09}
sample. In Fig.\ \ref{fig:kennicutt} we then plot the evolution of this
model, as well as the loci of the clouds from the \citet{Evans+09} and
\citet{Heiderman+10}, adding an upwards-pointing arrow to the latter
points, of length corresponding to one order of magnitude, to indicate
the likely underestimation of the SFR for massive clumps by the latter
authors.  We also plot the data from OMC-1 \citep[from ][]{VS+09} and
Taurus \citep[see e.g., ][]{Heiderman+10}. The model evolves from low to
high values of both $\Sigma_{\rm gas}$ and $\Sigma_{\rm SFR}$.

It is interesting to note that the model passes first through the locus
of the low-mass star-forming clouds and later near the locus of the
clumps forming massive stars. This means that the model predicts that
present-day, relatively quiescent, low-mass-star forming clouds may
evolve into massive-star-forming ones in a few to several Myr. A similar
conclusion was reached through numerical simulations by \citet{VS+09}.
This reinforces the idea that the dispersion of the observational data
is due to different evolutionary states of the clouds in a sample \citep[see,
e.g., ][]{Bigiel+10}.

\subsection{Stellar age distribution} \label{sec:age_dist}

One important prediction of our model is that the SFR increases over
time. This is because, as the cloud contracts and its mean density
increases, the fraction of star-forming gas in the cloud increases. An
increasing SFR has already been proposed by \citet{PS99, PS00, PS02}
on the basis of the age distribution in various low- and high-mass
clusters. However, this result has been questioned, since there is
evidence suggesting that the older stars are not genuine members of the
clusters, but rather belong to a different population \citep{Hartmann03,
BH07, HH08}.  Moreover, \citet{Hartmann03} has posed the conundrum that,
if most clouds form at an accelerated pace only over the last few Myr,
and form stars at a very slow rate over the previous 10 Myr or so, most
clouds should be found to be in the slow-star-forming period, but
this is not what is observed. Our evolutionary scenario for clouds may
offer a solution to this debate.

\begin{figure}[!ht] 
\plottwo{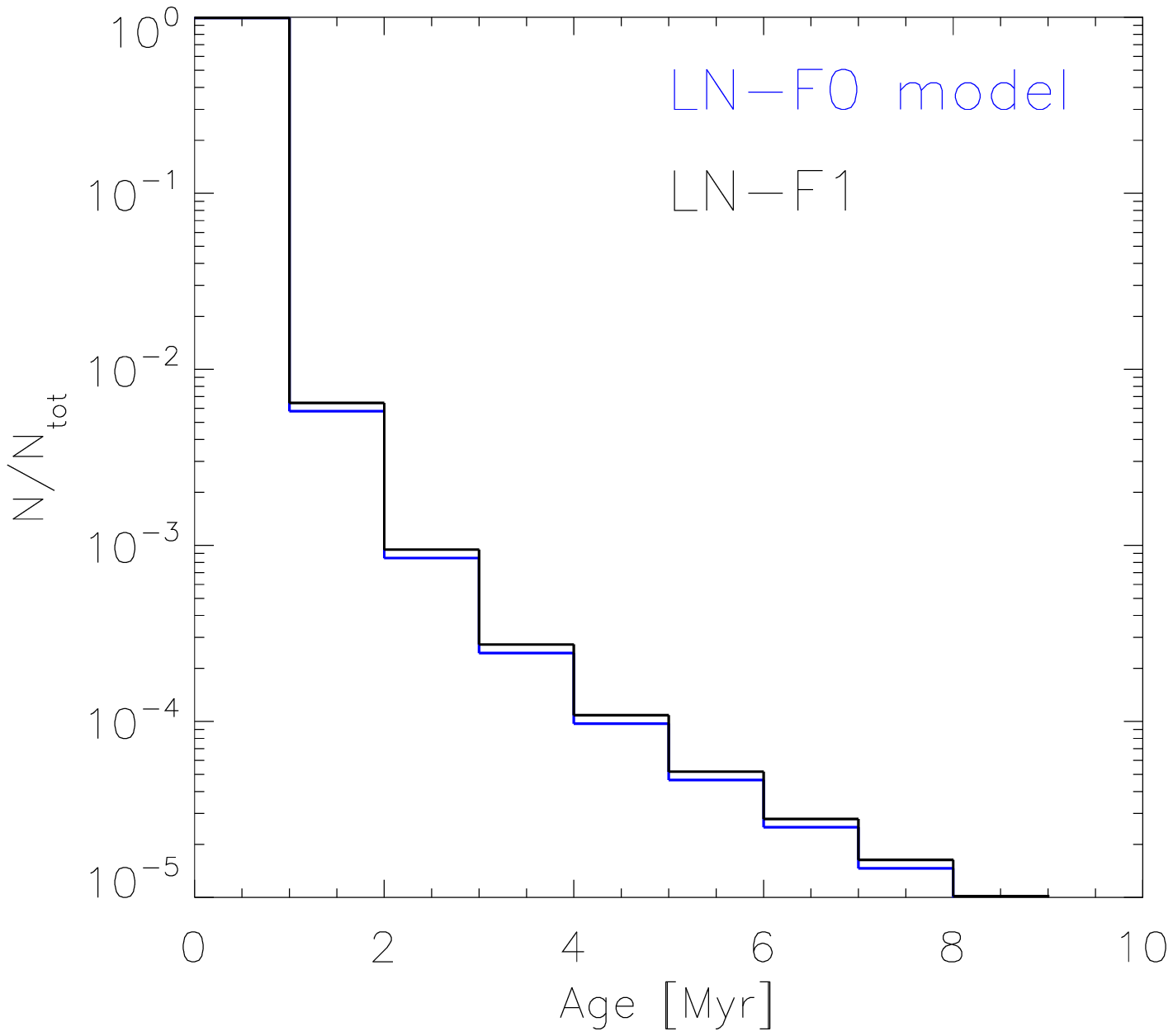}{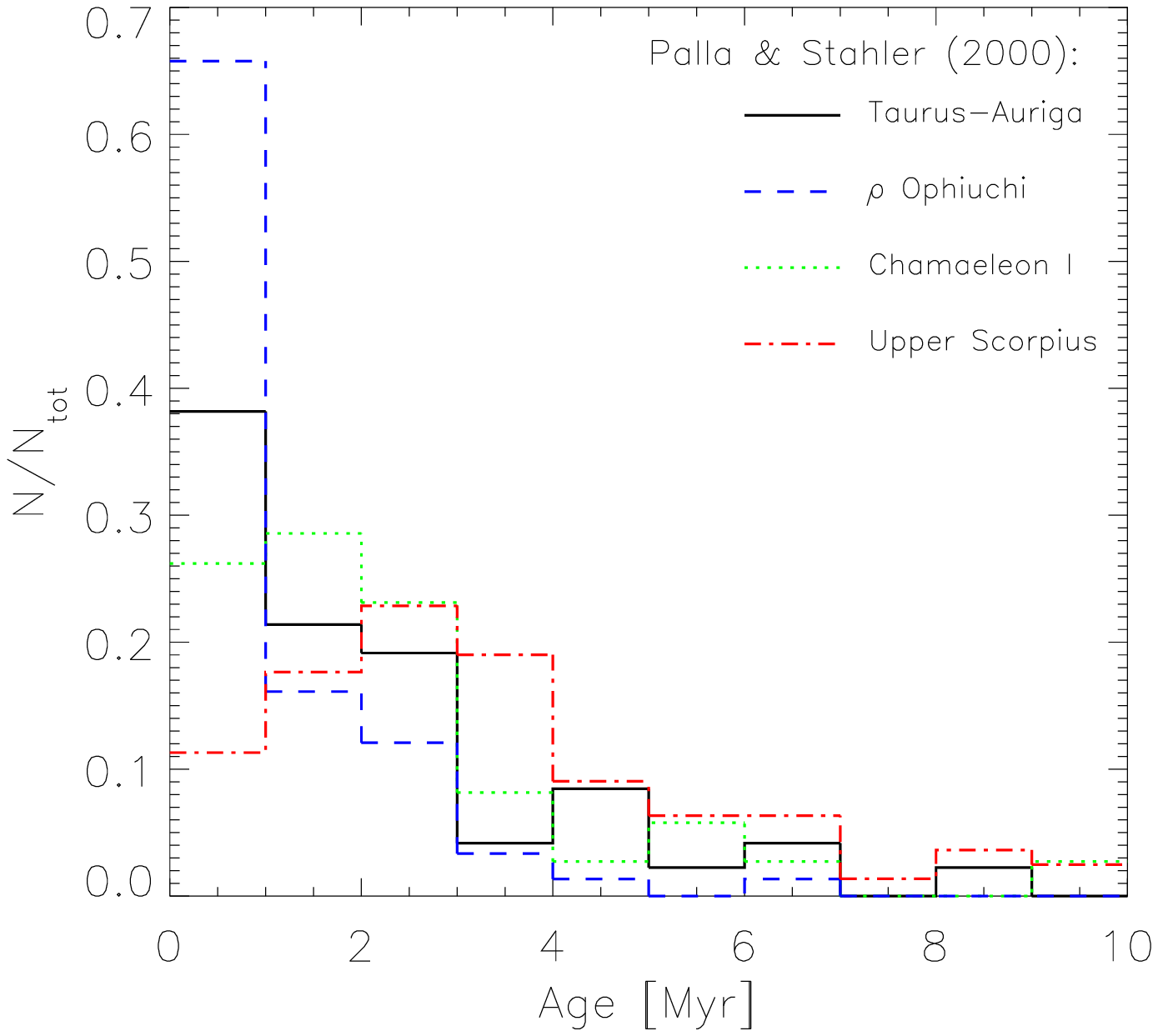}
\caption{{\label{fig:final_age_dist} {\it Left:} Stellar age distribution
for our calibrated model with $\Rinf = 10$ pc ($\Mmax  \approx 2000
\Msun $), calculated at the 
end of the cloud's evolution. {\it Right:} Compilation of the age
histograms for the associations studied by \citet{PS00}. }}
\end{figure}

Because in our model we compute the total mass of stars ($\Delta M$)
formed at each time step, we can readily obtain the total number of
stars formed during that time step as the integral over all masses of
the IMF, normalized to $\Delta M$. The {\it left panel} of Fig.\
\ref{fig:final_age_dist} shows the stellar age histogram for our
calibrated model, with $\Rinf = 10$ pc ($\Mmax \approx 2000 \Msun$) at
the end of its 
life -- i.e., when it has completely lost its gas. We show the
histograms for a case with feedback off (model LN-F0) and one with
feedback on (model LN-F1). It is clear from this
figure that indeed the age distribution is concentrated towards young
ages, although a few older stars exist.

These results can be compared with the age histograms presented by
\citet{PS00} for the Orion Nebula Cluster (ONC), the Taurus-Auriga
region, Lupus, $\rho$-Oph, Chameleon, Upper Scorpius, and IC348. In
these clusters, the fraction of stars with ages up to 1 Myr ranges from
$\sim 30$\% to 66\%, while the fraction of stars with ages up to 4 Myr
is in the range 80 -- 97\%. Moreover, only in the case of Upper Scorpius
does the age histogram peak at an age larger than 1 Myr, namely at 3
Myr.  For this association, the fraction of stars with ages $\le 1$ Myr
is only 11\%, while the fraction with ages $\le 4$ Myr is 71\%. We show
a compilation of these in the {\it right panel} of Fig.\
\ref{fig:final_age_dist}.

We can see that, qualitatively, the stellar age histogram at the end of
our model's life resembles those of \citet{PS00}, although,
quantitatively, the model's histogram in Fig.\ \ref{fig:final_age_dist}
is much more concentrated towards short ages. However, this must be
due to the fact that it was calculated at the end of the model's
evolution. Clearly this is not the case for the clusters and groups
analyzed by \citet{PS00}, because, as those authors themselves point
out, in most cases the clusters are still embedded in their parent
clouds, with only Upper Scorpius being already exposed. This means that
we should consider our model {\it before} the end of its life.

\begin{figure}[!ht] 
\begin{centering}
\epsfig{file=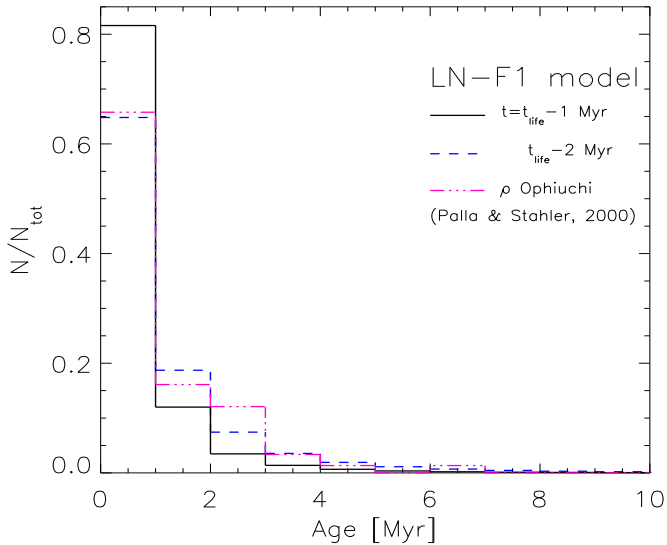,height=8cm,width=10cm}
\caption{{\label{fig:age_dist_vs_t} Stellar age distribution
for our calibrated model with $\Rinf = 10$ pc ($\Mmax  \approx 2000
\Msun $), calculated 
at one and two Myr before the end of the cloud's evolution, compared
with the corresponding distribution for the $\rho$-Oph association
\citep{PS00}. }}
\end{centering}
\end{figure}

In Fig.\ \ref{fig:age_dist_vs_t} we show the age histogram for the
calibrated model, calculated at 1 and 2 Myr before the end of its
evolution, and compare it with one of the histograms from \citet{PS00}
-- that for $\rho$-Oph. We see that the histogram becomes less peaked as
earlier times before the cloud's destruction are taken, becoming more
closely resemblant to the histograms of \citet{PS00}. This is because,
as the SFR increases towards later times, the fraction of young stars
becomes increasingly larger. In particular, the histograms for 1 and 2
Myr before the cloud's dispersal seem to bracket the histogram for
$\rho$-Oph. As seen from the {\it right panel} of Fig.\
\ref{fig:final_age_dist}, the other regions are less concentrated
towards short ages. According to our model, then, $\rho$ Oph is a
somewhat more evolved region, well matched by our model at $\lesssim 2$
Myr before its dispersal, while the other regions would correspond to
somewhat less evolved stages, at slightly earlier times (3--4 Myr)
before destruction.

These results suggest a possible resolution of the debate between the
Palla-Stahler and the Hartmann groups. Specifically, although our model
indeed predicts an increase of the SFR in collapsing clouds, this
does not conflict with the conundrum posed by \citet{Hartmann03}: no
fully formed molecular clouds are observed without significant amounts
of star formation {\it because the clouds themselves are evolving.}
Thus, at the time when they had much lower SFRs, they were not fully
formed yet, and thus not identifiable as large molecular clouds. Indeed,
the clouds' mean density was lower, and thus, in reality, they probably
consisted of a few molecular clumps immersed in a still-atomic
interclump medium. Only in the last few Myr of their evolution, the
clouds are dense enough on average that most of their bulk is already
molecular, and by that time they are forming stars at a much higher
rate, as observed. A similar conclusion has been recently reached on the
basis of numerical simulations by \citet{HBH12}.

\section{Discussion} \label{sec:discussion}

\subsection{The constant-$\Mrms$ assumption} \label{sec:cst_Mach}

A feature of our model that may appear odd at first sight is that we
have taken $\mathcal{M}_{\rm rms}=3$ as the (constant) fiducial value
for the rms Mach number of the turbulence within the cloud, as it is
contrary, for example, to the famous \citet{Larson81} velocity
dispersion-size scaling relation. However, it must be recalled that in
this paper we are specifically assuming that such a relation, or its
more modern rendition by \citet{Heyer+09}, is a manifestation of the
gravitational contraction of the cloud, rather than a feature of the turbulence
\citep{BP+11}. Thus, the relevant rms Mach number must be the remainder
after the collapsing motions have been removed. A competition may be set
up between the transfer of kinetic energy from the collapsing motions to
the turbulent ones and the dissipation \citep{VS+98, KH10}, and so, in
the absence of a reliable model, we consider that the assumption of a
constant rms Mach number with the value typical for the CNM \citep{HT03}
is reasonable, although a possible alternative recipe for its initial
value would be to take it equal to the inflow Mach number
\citep{Banerjee+09}. We consider that further work is necessary to
better constrain this parameter.

\subsection{The lognormal PDF assumption} \label{sec:logn_PDF}

A similar situation arises for the density PDF in the cloud, which we
have assumed to have a lognormal shape, even though it is well known
that star-forming clouds develop a power-law tail at high densities
\citep{Klessen00, DB05, VS+08, Kainulainen+09, Kritsuk+11,
BP+11b}. However, \citet{Kritsuk+11} have suggested that such power-law
tails are the effect of the development of local collapsing sites with
power-law density profiles. In this case, as explained in \S
\ref{subsec:PDF}, the power-law tail in the PDF would be the {\it
result} of the collapse, rather than the {\it seed} for it, and thus the
relevant PDF for the seeds for future collapse should be the underlying
lognormal one, after removal of the already-collapsing regions.
Moreover, there is no complete theory for how the density PDF should
evolve in time from a lognormal to a power-law. Unknowns such as the
timescale for the transition, the density at which the power-law tail
starts, and the final slope of this region are uncertain at present. As
above, we consider that further work is necessary to clearly resolve
this issue, and in the meantime we settle for the lognormal PDF
assumption.

\subsection{Accelerating star formation} \label{sec:accel_SF}

An important precision is in order concerning the acceleration of star
formation in our model. Indeed, our model predicts that the star
formation {\it rate} increases over time, and therefore, star formation
(SF; strictly speaking, the instantaneous stellar mass, $\Ms(t)$) {\it
accelerates}. However, it is common to find the statement in 
the literature that it is the SFR that accelerates. This is {\it not}
the case for our model with feedback. The SFR is the time
derivative of $\Ms$. Since the SFR increases in time, the second time
derivative of $\Ms$ is positive, and thus the SF accelerates. However,
the second time derivative of the SFR (the third derivative of $\Ms$)
is negative for our model with feedback (see the {\it left panel} of
Fig.\ \ref{fig:SFR_SFE_ne6}), and thus strictly speaking the SFR
decelerates.

\subsection{Room for improvement} \label{sec:lims}

In the present model, we have bypassed the supporting effect of all
forms of pressure, and replaced it by the empirical ``Larson factor'',
$\fL$, which effectively lengthens the timescale for collapse. In the
case of the original work by \citet[][ Appendix C]{Larson69}, this
factor represented the support from thermal pressure which,
incidentally, should be most important during the earlier stages of the
collapse. Calibrating against the SAF1 simulation by \citet{VS+10}, we
found a value of $\fL$ rouhly 8\% larger than the one found by Larson,
suggesting perhaps that turbulent pressure added a certain (small)
amount of support (the magnetic field was not included in that
simulation). Including physically-motivated terms into the collapse
prescription that account for thermal, turbulent and magnetic support is
an important goal, which we will attempt to pursue in a future
contribution.

Nevertheless, it is interesting that our model, calibrated in the
non-magnetic case of the SAF1 simulation, gives a good match to a number
of observational properties of molecular clouds in a wide range of
masses. This suggests that magnetic support is not crucial in these
objects. In turn, this is consistent with the recent realization that
star-forming clouds tend to be magnetically supercritical in general
\citep{Bourke+01, Crutcher+03, TC08}, and thus they should be
essentially in a free-fall regime.

Finally, in this paper we have not considered the effect of supernova
explosions towards the late evolutionary stages of the clouds. This may
help in reducing the model's SFR at those stages, probably bringing it
to better agreement with the observations (cf.\ Fig.\
\ref{fig:kennicutt}).

\section{Summary and Conclusions} \label{sec:sum}

In this paper we have developed a semi-empirical analytical model
\citep[based on simulations by][] {VS+10} in which a MC is formed by
converging WNM flows. We assumed that the inflow collision produces a
CNM cloud, through nonlinear triggering of the thermal instability, and
that the cloud becomes turbulent through the combined action of the
latter and various other dynamical instabilities, such as the nonlinear
thin shell, Kelvin-Helmholtz, and Rayleigh-Taylor ones. We assumed that
the rms Mach number of this turbulence remains fixed at the typical
values in the CNM ($\mathcal{M}_{\rm s} \approx 3$), and that over its
evolution, the cloud develops further nonthermal motions related to its
collapse, not its internal turbulence. We also assumed that the cloud
forms stars with a \citet{Kroupa01}-type IMF, so that massive stars
only appear when a sufficiently large number of stars has formed to
adequately sample the high-mass tail of the IMF. Finally, we assumed
that the density PDF in the cloud has a lognormal shape and a fixed width
(corresponding to a constant turbulent Mach number $\Mrms$), but whose
maximum shifts towards higher densities as the cloud contracts and
becomes denser on average.

Using the same WNM inflow parameters as the simulation labeled SAF1 from
\citet{VS+10}, namely $R_{\rm inf}=64 \, {\rm pc}$, $n_{\rm inf}=1 ~{\rm
cm}^{-3}$, and $\mathcal{M}_{\rm rms}=3$, we calibrated the model by
searching the density threshold for star formation, $\nsf$, that best
matched the simulation's evolution of the SFR and SFE. Our calibrated
value was $n_{\rm SF}=10^6~ {\rm cm}^{-3}$. With the $\nsf$, $\ninf$,
and $\Mrms$ parameters fixed, the only remaining free parameter of the
model is the WNM inflow radius $\Rinf$, which essentially controls the
maximum mass reached by the model cloud, $\Mmax$. Varying this parameter
we then match the model to clouds of various masses, and compare with
various properties of such clouds.

The generic behavior of the model cloud, with the parameters of the SAF1
simulation, are as follows: i) The size of the model cloud decreases
faster (by gravitational contraction) in a case {\it without} stellar
feedback (model LN-F0) than in a case {\it with} it (model LN-F1). This
is because feedback partially evaporates the cloud, thus reducing its
gravitational potential, and slowing its collapse.  ii) The model
without feedback approaches a final state of zero size with finite mass
(a singularity), while the case with feedback approaches a final state
of zero mass at finite size; i.e., it is completely consumed by SF and
ionization before if reaches zero size. iii) Although the SFR increases
in both cases, it {\it accelerates} over time in the LN-F0 model, while
it {\it decelerates} over time in the LN-F1 model.

We then set out to apply the model to explain a number of observed
features of molecular clouds of a wide range of masses. First, we
compared the predictions of our model with the evolutionary scenario for
GMCs recently proposed by \citet{Kawamura+09}, in which the GMCs start
out having no massive stars, then have reduced numbers of them, so as to
only have isolated HII regions, and finally have large numbers of them,
so as to clearly contain massive clusters. We find that our model, with
a value of $\Rinf$ that gives $\Mmax \sim 10^5~\Msun$, comparable to the
mass range reported by those authors, spends similar times in each of
the stages reported by them. 

We also investigated a model cloud with $\Rinf = 10$ pc, corresponding
to $\Mmax \sim 2000~\Msun$. We find that such a model cloud evolves in
the $\SSFR$-$\Sg$, or Kennicutt-Schmidt, diagram, in such a way that it
passes first through the locus of individual low-mass-star forming
clouds and later through the locus of high-mass-star forming clumps, as
reported by \citet{Evans+09} and \citet{Heiderman+10}. Next, we compared
an evolved stage of the calibrated model, also using $\Rinf = 10$ pc,
with the physical conditions in the OMC-1 massive clump, finding that it
has similar physical conditions after $\sim 26$ Myr of evolution since
its parent molecular cloud first formed, although it spends only about
2 Myr in a state comparable to OMC-1.

Finally, we investigated the stellar age distribution in our
isolated-cloud model with $\Rinf = 10$ pc, showing that, taken a few Myr
before the end of the cloud's life, it is consistent with the
corresponding distributions presented by \citet{PS00} for various
clusters and associations. Furthermore, the model predicts that the
shape of this age distribution depends on the evolutionary stage of the
system, being more peaked towards young ages as the system grows older,
because of its increasing SFR.

We conclude that our evolutionary and collapsing model of molecular
clouds adequately represents actual clouds of a wide range of masses,
with no need whatsoever for the consideration of equilibrium states. In
this sense, the present model, although idealized, represents a
promising first attempt at a non-equilibrium model for molecular clouds
and their star-forming properties.

\acknowledgements

We acknowledge helpful discussions with Javier Ballesteros-Paredes and
Jos\'e Franco. An anonymous referee suggested to perform the comparison
against the SAF1 simulation, helping us to better constrain the
model. This work has received partial finacial support from CONACYT
grant 102488

\clearpage

\end{document}